\documentclass[sigconf]{acmart}

\AtBeginDocument{%
  \providecommand\BibTeX{{%
    \normalfont B\kern-0.5em{\scshape i\kern-0.25em b}\kern-0.8em\TeX}}}

\copyrightyear{2024} 
\acmYear{2024} 
\setcopyright{acmlicensed}\acmConference[ASE '24]{39th IEEE/ACM
International Conference on Automated Software Engineering }{October
27-November 1, 2024}{Sacramento, CA, USA}
\acmBooktitle{39th IEEE/ACM International Conference on Automated Software
Engineering (ASE '24), October 27-November 1, 2024, Sacramento, CA, USA}
\acmDOI{10.1145/3691620.3695020}
\acmISBN{979-8-4007-1248-7/24/10}

\usepackage{multirow,makecell}
\usepackage{tcolorbox}
\usepackage{color,xcolor}
\usepackage{listings,amsfonts}
\usepackage{caption}
\usepackage{subcaption}
\usepackage{threeparttable}
\usepackage{bbding}
\usepackage{graphicx}
\usepackage{booktabs} 
\usepackage{longtable}
\usepackage{pifont}
\usepackage[linesnumbered,ruled,vlined]{algorithm2e}

\definecolor{customblue}{HTML}{006ca6}
\definecolor{customgreen}{HTML}{009264}
\definecolor{custombrown}{HTML}{ff3d00}
\AtEndPreamble{
 \usepackage{hyperref}
 \hypersetup{
  colorlinks = true,
  linkcolor = customblue,
  anchorcolor = customblue,
  citecolor = customgreen,
  filecolor = customblue,
  urlcolor = customblue
 }
}

\newcommand{\tool}[1]{\textsc{WaDec}}

\newcommand{\ea}{\textit{et al. }}

\title{\tool{}: Decompiling WebAssembly Using Large Language Model}

\begin{document}

\author{Xinyu She}
\authornote{Xinyu She and Yanjie Zhao contributed equally to this work.}
\authornote{The full name of the authors' afiliation is Hubei Key Laboratory of Distributed System Security, Hubei Engineering Research Center on Big Data Security, School of Cyber Science and Engineering, Huazhong University of Science and Technology.}
\orcid{https://orcid.org/0009-0001-2988-7042}
\affiliation{%
  \institution{Huazhong University of Science and Technology}
  \city{Wuhan}
  \country{China}
}
\email{xinyushe@hust.edu.cn}

\author{Yanjie Zhao}
\orcid{https://orcid.org/0000-0001-8793-5367}
\authornotemark[1]
\authornotemark[2]
\affiliation{%
  \institution{Huazhong University of Science and Technology}
  \city{Wuhan}
  \country{China}
}
\email{Yanjie_Zhao@hust.edu.cn}

\author{Haoyu Wang}
\authornote{Haoyu Wang is the corresponding author (haoyuwang@hust.edu.cn).}
\authornotemark[2]
\orcid{https://orcid.org/0000-0003-1100-8633}
\affiliation{%
  \institution{Huazhong University of Science and Technology}
  \city{Wuhan}
  \country{China}
}
\email{haoyuwang@hust.edu.cn}

\begin{abstract}
WebAssembly (abbreviated Wasm) has emerged as a cornerstone of web development, offering a compact binary format that allows high-performance applications to run at near-native speeds in web browsers. Despite its advantages, Wasm's binary nature presents significant challenges for developers and researchers, particularly regarding readability when debugging or analyzing web applications. Therefore, effective decompilation becomes crucial.
Unfortunately, traditional decompilers often struggle with producing readable outputs. While some large language model (LLM)-based decompilers have shown good compatibility with general binary files, they still face specific challenges when dealing with Wasm.

In this paper, we introduce a novel approach, \tool{}, 
which is the first use of a fine-tuned LLM to interpret and decompile Wasm binary code into a higher-level, more comprehensible source code representation. The LLM was meticulously fine-tuned using a specialized dataset of \texttt{wat-c} code snippets, employing self-supervised learning techniques. This enables \tool{} to effectively decompile not only complete \texttt{wat} functions but also finer-grained \texttt{wat} code snippets.
Our experiments demonstrate that \tool{} markedly outperforms current state-of-the-art tools, offering substantial improvements across several metrics. It achieves a code inflation rate of only 3.34\%, a dramatic 97\% reduction compared to the state-of-the-art's 116.94\%. Unlike the output of baselines that cannot be directly compiled or executed, \tool{} maintains a recompilability rate of 52.11\%, a re-execution rate of 43.55\%, and an output consistency of 27.15\%. Additionally, it significantly exceeds state-of-the-art performance in AST edit distance similarity by 185\%, cyclomatic complexity by 8\%, and cosine similarity by 41\%, achieving an average code similarity above 50\%.
In summary, \tool{} enhances understanding of the code's structure and execution flow, facilitating automated code analysis, optimization, and security auditing.

\end{abstract}

\maketitle

\section{Introduction}
\label{sec:introduction}

WebAssembly (abbreviated Wasm) emerges as a pioneering open standard that redefines the landscape of cross-platform computing~\cite{wasm_web}. It specifies a portable, language-agnostic, and architecture-independent binary instruction format engineered to deliver near-native execution speeds across a broad spectrum of platforms, with a particular emphasis on web browsers. Widely adopted in various fields, Wasm has seen extensive applications in fields such as blockchain~\cite{bian2019poster,protzenko2019formally}, IoT~\cite{li2021wiprog}, and cloud computing~\cite{menetrey2022webassembly}. 

The widespread adoption of Wasm has increased the demand for in-depth analysis, often relying on decompilers to reverse-engineer its binary format.
Decompilers are transformative tools that convert low-level machine language into high-level programming languages, playing a pivotal role in fields such as binary program comprehension~\cite{park2023static}, reverse engineering~\cite{yu2023cfg2vec}, legacy code migration~\cite{liu2023decompiling}, and software security analysis~\cite{7546501}. By converting executable code into a human-readable format, they facilitate code analysis, comprehension, and reengineering. High-quality decompilation is thus essential for developers to manage complex binary programs, reducing maintenance and refactoring costs.

Traditional decompilation frameworks like \texttt{Ghidra} offer preliminary support for Wasm files but face significant limitations cause they are not designed for Wasm originally. Tools specifically designed for Wasm, such as \texttt{Wasm2c}~\cite{wasm2c} and the \texttt{Wasm-decompile}~\cite{Wasm-decompile} in Wabt~\cite{Wabt}, also encounter severe challenges. Specifically, the \texttt{C} code generated by \texttt{Wasm2c} often suffers from readability and structuring issues, along with significant code bloat. Similarly, while \texttt{Wasm-decompile}r uses a \texttt{C}-like representation, its generated code is difficult to adopt widely due to the lack of standardization and accuracy in capturing Wasm semantics.

Recent advancements in large language models (LLMs), especially code LLMs like CodeLlama~\cite{2024codellama}, have demonstrated their exceptional abilities in code-related tasks~\cite{hou2024large}. Their capacity to understand complex semantics and structural nuances makes them adept at tasks requiring deep insights into program semantics and structures. These developments have paved the way for producing high-quality, intelligible decompiled outputs.
In decompilation, Tan~\ea~\cite{tan2024llm4decompile} employed an LLM, DeepSeek~\cite{guo2024deepseek}, for assembly code decompilation, improving accuracy and readability.
Cao~\ea~\cite{Cao_Liang_Chen_Hu_2022} introduced NeurDP, a neural networks-based method, using LLVM IR to bridge low-level and high-level programming languages.

\begin{table*}[t]
\centering
    \caption{The comparison between our \tool{} approach and the state-of-the-art methods.}
    \label{tab:comparison}
      \begin{threeparttable}
  \resizebox{0.9\linewidth}{!}{
    \begin{tabular}{|c|l|ccccc|}
    \hline
    \multicolumn{2}{|c|}{\multirow{2}{*}{\textbf{Decompilation-related tools}}} & \multicolumn{5}{c|}{\textbf{Challenges}} \\ 
    \cline{3-7}    
    \multicolumn{2}{|c|}{} & {Source code generation} & {Good readability} & {Code bloat resolution} & {Fine-grained snippet\tnote{\textit{a}}}\hspace{0.3em} handling & {Nested loop handling} \\
    \hline
    \multirow{3}{*}{Traditional tools for Wasm} & Ghidra~\cite{Ghidra} & \ding{51}     & \ding{55}     & \ding{55}     & \ding{55}     & \ding{51} \\
\cline{2-7}          & Wasm2c\cite{wasm2c} & \ding{51}     & \ding{55}     & \ding{55}     & \ding{55}     & \ding{51} \\
\cline{2-7}          & Wasm-decompile~\cite{Wasm-decompile} & \ding{51}     & \ding{55}     & \ding{55}     & \ding{55}     & \ding{51} \\
    \hline
    \multirow{2}{*}{LLM-based tools for general binary} & LLM4Decompile~\cite{tan2024llm4decompile} & \ding{51}     & \ding{51}     & \ding{51}     & \ding{55}     & \ding{55} \\     
\cline{2-7} & NeurDP~\cite{Cao_Liang_Chen_Hu_2022} & \ding{51}     & \ding{55}     & \ding{55}     & \ding{51}     & \ding{55} \\ 
    \hline
    \multirow{3}{*}{LLM-based tools for Wasm} & WASPur~\cite{romano2023automated} & \ding{55}     & \ding{55}     & \ding{55}     & \ding{55}     & \ding{55} \\
\cline{2-7}          &  SnowWhite~\cite{Lehmann_Pradel_2022} & \ding{55}     & \ding{55}     & \ding{55}     & \ding{55}     & \ding{55} \\
\cline{2-7}          &  WasmRev~\cite{huang2024multi} & \ding{55}     & \ding{55}     & \ding{55}     & \ding{55}     & \ding{55} \\
    \hline
    Our approach & \textbf{\tool{}} & \ding{51}     & \ding{51}     & \ding{51}     & \ding{51}     & \ding{51} \\
    \hline
    \end{tabular}}
\begin{tablenotes}
\footnotesize
\item[\textit{a}] This refers to code snippets that are more fine-grained than functions.
\end{tablenotes}
  \end{threeparttable}
\end{table*}

Unfortunately, \textbf{existing LLM-based decompilation tools for general binaries face unique challenges when handling Wasm}. 
Unlike traditional binaries, Wasm's lack of support for complex data structures, stack-based architecture, and linear memory model, enhance performance but reduce readability. Its structured block-based control flow and the storage of global variables and string constants in data segments present additional obstacles. For instance, applying Cao~\ea~\cite{Cao_Liang_Chen_Hu_2022}'s NeurDP to Wasm disrupts the clear correspondences between \texttt{C} and Wasm code due to differences in the number of blocks between Wasm and LLVM IR. Additionally, Wasm's verbose text representation, often thousands of lines long, further complicates decompilation, leading to a significant drop in quality when using existing LLM-based methods.

There are also LLM-based efforts for Wasm decompilation. However, these works have generally \textbf{avoided the primary goal of producing high-level, readable source code}, instead focusing on subtasks such as code summarization, type recovery, function identification, etc. For example, Romano~\ea~\cite{romano2023automated} developed WASPur, an LLM-based tool that predicts the purpose of Wasm functions by comparing them to known named functions. Similarly, Lehmann~\ea~\cite{Lehmann_Pradel_2022} proposed SnowWhite, a tool for recovering the precise, high-level parameters and return types of Wasm functions.
Huang~\ea~\cite{huang2024multi} created WasmRev, a tool trained using multimedia inputs and fine-tuned for type recovery, function intent recognition, and code summarization.

As illustrated in \autoref{tab:comparison}, the aforementioned decompilation tools struggle with poor readability, code bloat, lacking fine-grained support, etc.
To fill this gap, we introduce a novel approach, \tool{}, 
which utilizes a fine-tuned LLM to interpret and decompile Wasm binary code into a more comprehensible, higher-level source code representation. Initially, we convert the input Wasm binary files into a text format (i.e., \texttt{wat}~\cite{Wat}) that is easier to analyze. Given the unique code structure characteristics of Wasm, which include excessively long lines and string constants isolated from the function body, it is necessary to reduce the decompilation complexity. 
To achieve this, we segment the function code within the \texttt{wat} file according to Wasm's loop block structure. This slicing strategy allows for the individual decompilation of each block, streamlining the process and enhancing the manageability of the code.
To mitigate the variable naming confusion that often arises from block-wise decompilation, we implement a unified variable renaming scheme that ensures consistency across the decompiled source code. Finally, the decompiled code blocks are reassembled to produce the final output. Our strategies for block-wise decompilation and unified variable naming effectively address the unique structural challenges of Wasm code, resulting in a more readable and understandable decompiled output.

In summary, our key contributions are as follows:
\begin{itemize}
    \item We have contributed a valuable dataset to the community, consisting of 110,091 pairs of \texttt{wat} and \texttt{C} code snippets that are more granular than complete functions. The variables in the \texttt{C} code have been systematically renamed according to our variable renaming scheme. This dataset is expected to facilitate advancements in the research and development of Wasm decompilation tools, potentially enhancing their precision and efficiency.
    \item We introduce \tool{}\footnote{The replication artifact, including our dataset, is publicly available at 
    \url{https://github.com/XinyuShe/WaDec}.
    }, an approach that fine-tunes a pre-trained code LLM to process extensive Wasm files, producing source code-level decompiled outputs. To the best of our knowledge, \tool{} is the first initiative to fine-tune an LLM specifically for Wasm decompilation, achieving high-fidelity source code-level results. 
    \item We conducted an extensive experimental evaluation of \tool{}, comparing its performance with state-of-the-art Wasm decompilation tools. The results reveal that \tool{} surpasses other tools in both accuracy and readability. Notably, it maintains a low code inflation rate of 3.34\%, a 97\% reduction compared to the previous 116.94\%. \tool{} ensures that 52.11\% of outputs are directly re-compilable and achieves a re-execution consistency of 27.15\%. It also exceeds state-of-the-art performance in AST edit distance similarity by 185\%, cyclomatic complexity by 8\%, and cosine similarity by 41\%, maintaining an average code similarity above 50\%.
\end{itemize}

\section{Background}
\label{sec:background}
We now present the relevant background knowledge on Wasm, accompanied by an example to illustrate the challenges and limits of the state-of-the-art methods encountered in Wasm decompilation.

\subsection{Wat}
\label{sec:Wat}

\begin{figure*}[t]
    \centering
    \includegraphics[width=0.95\linewidth]{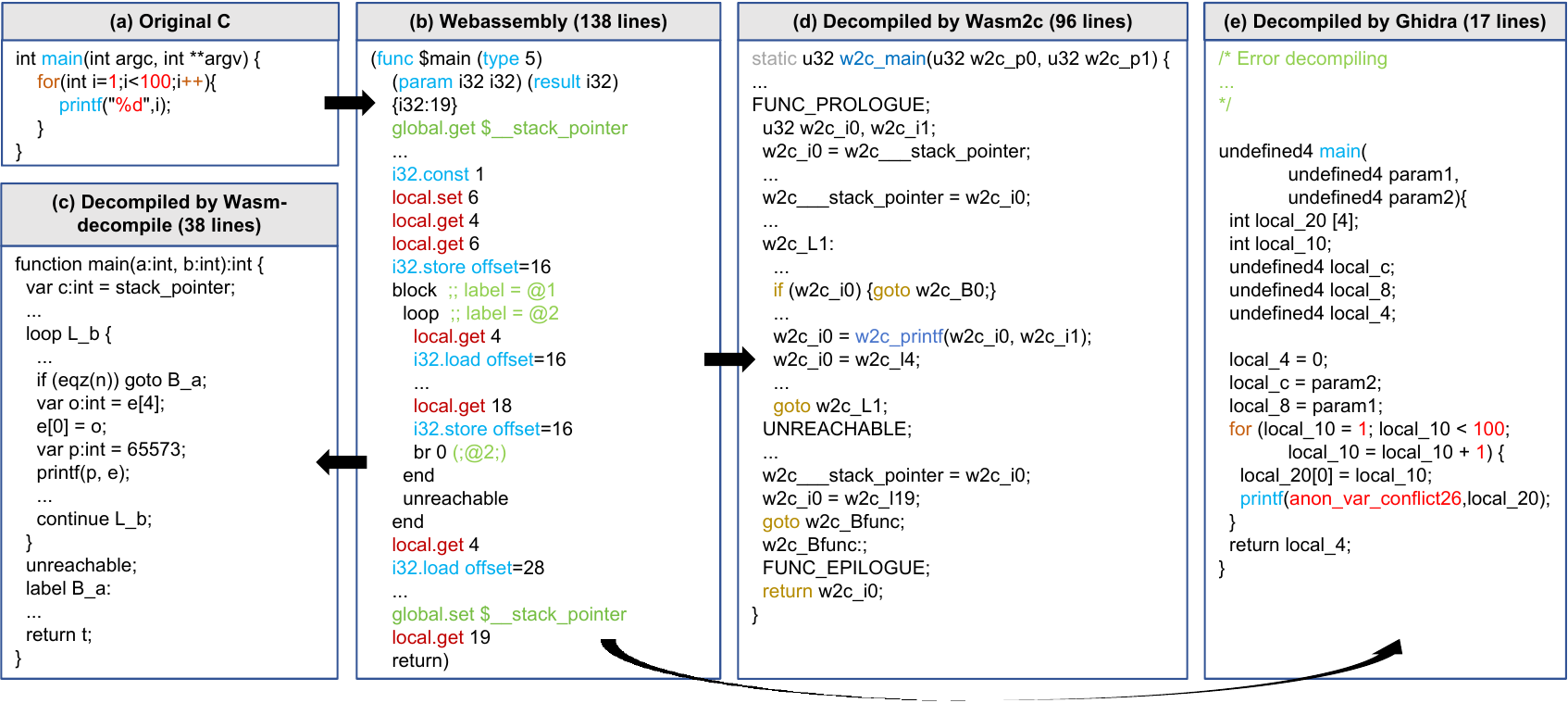}
    \caption{An example of the limitations of \texttt{Wasm2c}, \texttt{Wasm-decompile}, and \texttt{Ghidra}: (a) Original \texttt{C} code. (b) The compiled Wasm code in text format (i.e., \texttt{wat}). (c) The decompiled \texttt{C}-like syntax from Wasm by \texttt{Wasm-decompile}. (d) The decompiled result by \texttt{Wasm2c}. (e) The decompiled result by \texttt{Ghidra}.}
    \label{fig:state_of_art_example}
\end{figure*}

To improve readability, Wasm offers a textual format, \texttt{wat}, that clearly outlines the module's internal structures like types, memory, and function definitions. Both the textual (\texttt{wat}) and binary (Wasm) formats are theoretically equivalent, ensuring no information is lost during conversion~\cite{Wat}.
Therefore, \texttt{wat} can serve as a substitute for Wasm when readability and comprehension are required, as it is more easily understood and interpreted by both humans and LLMs without specialized knowledge of Wasm.
An example is illustrated in \autoref{fig:state_of_art_example}(a) and (b), which display a \texttt{C} source code example and its corresponding \texttt{wat} code. 

\subsection{Challenges in Wasm Decompilation}
\label{sec:challenges_in_Wasm_decompilation}
Wasm decompilation poses distinct challenges that are markedly different from those associated with traditional binary decompilation. These challenges can be categorized based on the approach taken, i.e., traditional static tools and machine learning (ML)-based methods. As shown in \autoref{fig:state_of_art_example}, we provide an example to illustrate the problems existing in traditional approaches.

\subsubsection{Traditional static tools}
Traditional static decompilers, such as \texttt{Ghidra}, \texttt{Wasm2c}, and \texttt{Wasm-decompile}, often struggle with the distinct characteristics of Wasm, resulting in several key issues:

\noindent\textbf{\underline{(C1) Poor readability.}} Decompiled code produced by traditional static tools often lacks the clarity and structured organization of the source code. Since Wasm is a low-level binary format, meaningful variable names and high-level constructs are lost during compilation, making the decompiled code difficult to interpret. \autoref{fig:state_of_art_example}(c), (d), and (e) show the decompilation results from \texttt{Wasm-decompile}, \texttt{Wasm2c}, and \texttt{Ghidra}, respectively. It is observed that there is a significant readability gap between their outputs and the source code. Especially for \texttt{Wasm2c} and \texttt{Wasm-decompile}, they essentially translate the code instruction by instruction, still preserving the Wasm instruction style, with hardly any improvement in readability.

\noindent\textbf{\underline{(C2) Significant code bloating.}} 
During the decompilation process, redundant intermediate representation may be retained or the code may be translated line by line, leading to bloated code that is cumbersome to work with and understand.
Due to space constraints, \autoref{fig:state_of_art_example}(b), (c), (d), and (e) do not display the complete code listings. The full line counts for the code are as follows: \texttt{wat} has 138 lines, \texttt{Wasm-decompile} has 38 lines, \texttt{Wasm2c} has 96 lines, and \texttt{Ghidra} has 17 lines.
Although the line counts of these three decompilation tools' outputs are reduced compared to the \texttt{wat}, they still exhibit significant code bloat compared to the source code. 
Moreover, the comparison is made at the function level. When comparing line counts at the file level, the bloat becomes even more staggering and user-unfriendly, as traditional decompilers do not distinguish between library functions and user-defined functions.

\noindent\textbf{\underline{(C3) Incapable of handling code snippets.}} Existing decompilation tools are primarily designed to handle complete programs or executable files. However, in many practical scenarios, developers often need to analyze and understand code snippets or individual functions rather than the entire program. 
Traditional decompilers struggle to handle such code snippets effectively, as they lack the context and metadata necessary to accurately reconstruct the semantics and data structures.
When presented with isolated code snippets, these decompilers may produce incomplete or erroneous outputs, as they cannot infer the missing information required for proper decompilation. This limitation hinders their applicability in situations where developers need to comprehend specific code sections, such as when reviewing code changes, analyzing security vulnerabilities, or performing code refactoring tasks.

\begin{figure*}[ht!]
    \centering
    \includegraphics[width=\linewidth]{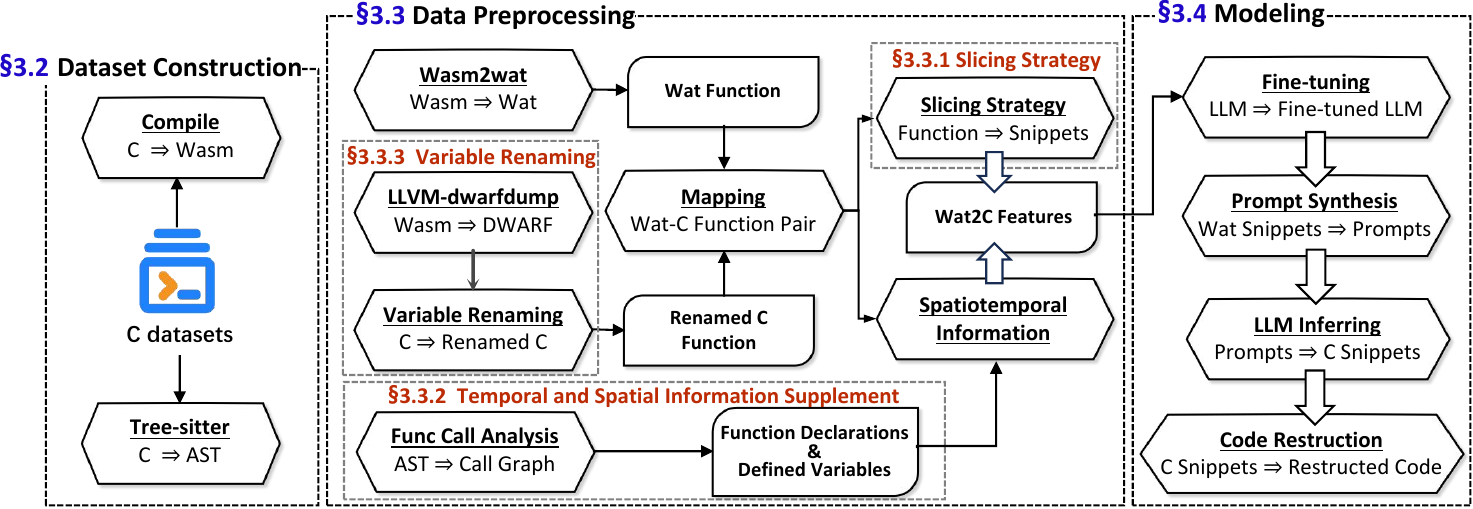}
    \caption{The workflow of \tool{}.}
    \label{fig:workflow}
\end{figure*}

\begin{figure*}[ht!]
    \centering
    \includegraphics[width=0.8\linewidth]{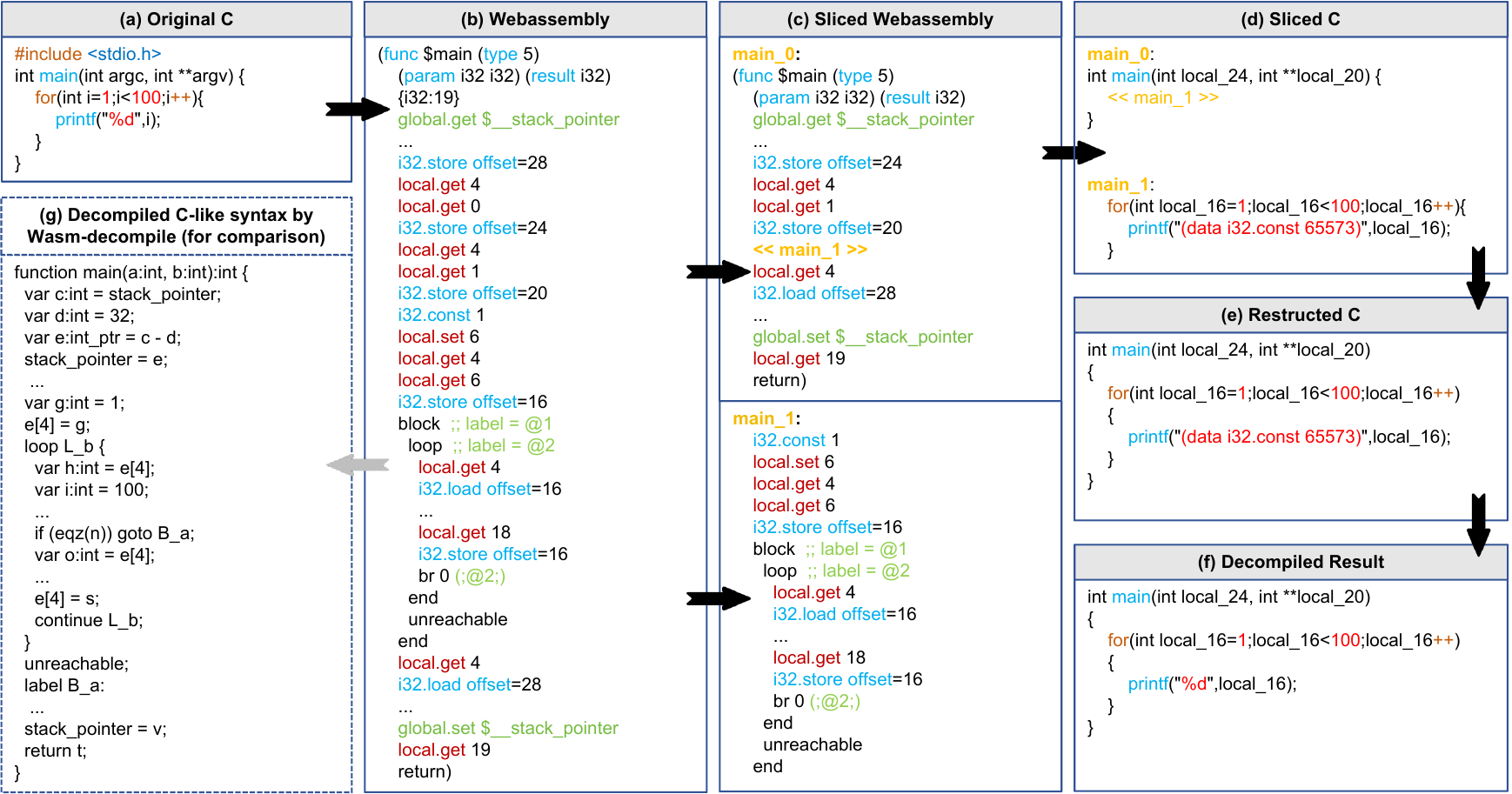}
    \caption{An example of \tool{}'s decompilation results: (a) Source code. (b) The compiled Wasm in text format. (c) Sliced \texttt{wat} snippets according to loop. (d) Results of the reverse engineering task of \tool{}. (e) Restructuring C snippets according to markers. (f) Decompiled result after string recovery. (g) The decompiled \texttt{C}-like syntax from Wasm by \texttt{Wasm-decompile}.}
    \label{fig:decompile_example}
\end{figure*}

\textbf{Our solution.} 
To solve these problems, we leverage a code LLM to achieve source code-level decompilation results. Code LLMs, specifically designed for code-related tasks, not only inherit LLMs' natural language processing capabilities but also possess extensive knowledge of code~\cite{2024codellama,guo2024deepseekcoder,MLM}. On one hand, code LLMs can fully comprehend and apply code logic and rules, eliminating redundant intermediate representation, thereby reducing code bloat and enhancing readability. On the other hand, unlike traditional tools that require complete functions or even entire codebases for decompilation, code LLMs can work with more granular code snippets supplemented by necessary contextual information. This allows them to understand and decompile finer-grained code snippets.

\subsubsection{ML-based Methods}
Although ML-based methods offer promising advancements in decompilation, they still face challenges:

\noindent\textbf{\underline{(C4) Unstable response.}}
ML-based models are probabilistic, which leads to unstable responses, particularly as input complexity increases. However, Wasm snippets are much longer and more complex than typical outputs, spanning hundreds or thousands of instructions with intricate control flows and data dependencies. Directly inputting large Wasm snippets can overwhelm these models, resulting in unstable outputs or irrelevant responses.
Even with structural slicing techniques like sliding windows, there is a risk of disrupting Wasm's unique structural characteristics and introducing variable naming inconsistencies across decompiled snippets.

\noindent\textbf{\underline{(C5) Weak ability to handle nested loops.}} The ability to handle nested loops is notably limited in AI-based decompilers. Such loops, which are prevalent in advanced software, present considerable challenges in decompilation, impeding accurate reconstruction of the original code's logic and structure. The complexity of nested loops can result in decompiled code that fails to reflect the intricate control flow and dependencies, thus affecting the utility and precision of AI-driven decompilation efforts.

\textbf{Our Solution.}
To address C4 and C5, we consider the block structure of \texttt{wat}, noting that loops in \texttt{wat} have a specific structure. By segmenting the code based on loops, we ensure that each input contains at most one loop. This approach effectively reduces the complexity of the model input and resolves the challenge of handling deeply nested loops.

\section{Approach}
\label{sec:approach}

\subsection{Overview}

\autoref{fig:workflow} illustrates the overall architecture of our \tool{} approach. It includes three processes: dataset construction (\S\ref{sec:dataset_construction}), data preprocessing (\S\ref{sec:data_preprrocessing}), and modeling (\S\ref{sec:Modeling}).

In dataset construction, we provide a detailed description of the composition of our dataset and design key features that are crucial to our analysis. 
In the data preprocessing process, we propose slicing strategies to structurally slice the \texttt{wat} code into snippets (\S\ref{sec:slicing strategy}); provide temporal and spatial information to supplement the missing information in the sliced snippets, to achieve better decompilation results (\S\ref{sec:information supplement}); and present a variable renaming scheme to unify the naming of variables in the decompiled results, thereby perfectly restoring the source code (\S\ref{sec:rename_scheme}).
In the modeling process, we introduce the model architecture and provide a detailed description of the composition of the prompts.

We illustrate the entire decompilation process using an example shown in \autoref{fig:decompile_example}. 
As illustrated in \autoref{fig:decompile_example}(d), (e), and (f), we achieved near-perfect Wasm decompilation results. 
The final results exhibit minimal divergence from the original \texttt{C} code, with the primary discrepancies being some inconsistencies in variable naming and the omission of library references.
Compared to the \texttt{C}-like syntax generated by \texttt{Wasm-decompile}, as shown in \autoref{fig:decompile_example}(g), our approach exhibits minimal code inflation and superior readability.

\subsection{Dataset Construction}
\label{sec:dataset_construction}
We now describe how we built our dataset, including collection and the features of the final dataset obtained.

\subsubsection{\textbf{Dataset Collection}}
\label{sec:dataset collection}
Given that Wasm is a relatively new technology, there is limited research on using LLMs for its decompilation, and no specialized datasets are readily available for direct use, necessitating the creation of our own.
To construct our dataset, we selected 51,768 \texttt{C} programs from the \texttt{bigcode/the-stack-dedup} dataset~\cite{Kocetkov2022TheStack} hosted on Hugging Face~\cite{huggingface}, along with 377 \texttt{C} programs from \texttt{TheAlgorithms/C}~\cite{algorithm_c} on GitHub. 
These programs were then compiled into Wasm with \texttt{Emscripten}~\cite{emscripten}.

\subsubsection{\textbf{Datasets Features}}
\label{sec:dataset_features}

\begin{table}[ht!]
  \centering
  \caption{The features of the training dataset.}
   \resizebox{1\linewidth}{!}{
    \begin{tabular}{rl}
    \toprule
    \multicolumn{1}{r}{\textbf{Feature}} & \multicolumn{1}{l}{\textbf{Discription}} \\
    \midrule
    \textbf{Wat snippet} & Segmented based on loop blocks. \\
    \midrule
    \textbf{C snippet} & Segmented based on loop blocks, corresponding to wat snippet. \\
    \midrule
    \textbf{Spatial info} & Function declarations for called functions. \\
    \midrule
    \textbf{Temporal info} & Local variables already defined before current snippet. \\
    \midrule
    \textbf{Offset2string} & Mapping from offsets to string constants. \\
    \bottomrule
    \end{tabular}%
    }
  \label{tab:features}%
\end{table}%

As shown in \autoref{tab:features}, we define the necessary features for the fine-tuning dataset.
The \texttt{wat} snippets and \texttt{C} snippets are more granular code snippets than entire functions, usually containing no more than one loop statement each. 
Specifically, \texttt{C} snippets are processed such that all string constants are replaced with formatted strings generated from their offset addresses in the data segment, and this mapping is stored in the \texttt{Offset2string} feature. Temporal and spatial information serves as an additional supplement, providing essential details to enrich the \texttt{wat} snippets.

\subsection{Data Preprocessing}
\label{sec:data_preprrocessing}
To reduce complexity for extra long Wasm functions, prevent prompts from exceeding LLM input limits, and maintain integrity, we preprocessed the dataset in three steps, which are interconnected gears and inseparable. 
(1) First, given that the length of \texttt{wat} can be theoretically infinite, while the input strings for LLMs are finite, and considering that LLMs do not perform optimally when directly handling complex loops, we adopted a slicing strategy to reduce the length and complexity of a single input of \texttt{wat} code. 
(2) Second, the sliced \texttt{wat} snippets suffered from information loss, which would lead to unstable and ineffective results if directly fed into the model. Therefore, we augmented the prompts with the necessary spatial and temporal information to compensate for this loss. 
(3) Finally, although the result of each decompilation meets expectations, the variable names for the same variables varied across different decompilations, introducing randomness. To address this, we proposed a unified variable naming scheme that is independent of the decompilation order. This approach allows for the seamless integration of results from multiple decompilations, restoring the logic and semantics of the original source code.

\begin{algorithm}[t]
\caption{Program Slicing Strategy}
\label{algorithm:Program_Slicing_Strategy}
\SetKwFunction{FMain}{Slicing}
\SetKwProg{Fn}{Function}{:}{}
\Fn{\FMain{program}}{
    blocks $\gets$ \{ \} \\
    FuncIndexList $\gets 
    [0,1,2,\ldots, Program.FuncNum-1]$\\
    $OrderedFuncs$ $\gets$ [ ] \\
    \tcp{\footnotesize Order functions based on dependencies}
    \While{FuncIndexList}{
        \For{$i \in FuncIndexList$}{
            \tcp{\footnotesize Select the function whose all invocations are included in $OrderedFuncs$}
            \If{$FuncIdsCalledBy(i) \in OrderedFuncs$}{
                $OrderedFuncs$.append(i)\\
                FuncIndexList.remove(i)\\
            }
        }
    }
    \tcp{\footnotesize Slice the function by loop block}
    \For{$function \in OrderedFuncs$}{
        LoopBlocks $\gets$ GetAllLoops(function)\\
        \For{$(i,start,end) \in \text{enumerate}(LoopBlocks)$}{
            block\_id $\gets \{function.name\} \_ \{i\}$\\
            blocks[block\_id] $\gets$ Snippet(start,  end)\\
            \tcp{\footnotesize Replace the inner loop with $marker$}
            \For{$(j,st,ed) \in \text{enumerate}(LoopBlocks)$}{
                \If{j < i+1}{
                    continue
                }
                \If{$end < st$}{
                    break
                }
                inner\_block\_id $\gets \{function.name\}\_\{j\} $\\
                $marker$ $\gets$ $<<$ inner\_block\_id $>>$\\
                Replaceloop(block\_id, st, ed, $marker$)\\
            }
        }
    }
    \KwRet blocks\\
}
\end{algorithm}

\subsubsection{\textbf{Slicing Strategy}}
\label{sec:slicing strategy}

As demonstrated in \autoref{fig:decompile_example}(b) and (c), 
we showcase that \tool{} leverages the unique block structure inherent in Wasm to perform structured slicing of \texttt{wat} code.
Typically, these blocks enclose loops or conditional statements, forming the backbone of our slicing strategy. Our approach entails structural slicing aligned with loop boundaries, ensuring that each extracted code fragment contains no more than a single loop statement alongside an arbitrary assortment of conditional statements.

\underline{\textbf{Algorithm.}} We present a slicing algorithm, as shown in 
\autoref{algorithm:Program_Slicing_Strategy}, which is utilized in both the dataset construction and subsequent evaluation phases. The algorithm initiates by reordering all functions within the program based on their dependencies (lines 3–9), positioning the already-included functions at the forefront. This ordering ensures that the eventual decompiled source code is free from dependency issues. Starting from line 10, the algorithm processes each function individually. Our implementation is based on tree-sitter~\cite{Tree_sitter}, a robust parser that facilitates the syntactic analysis required for our structural slicing, utilizing both Clang-AST and \texttt{Wat}-AST parsers. The algorithm initially identifies all the start and end lines of loops within a function (line 11), sorting them by the start line number. 
For each loop block, subsequent loop blocks can either be nested within the current one or positioned entirely after the current loop block. 
Hence, the algorithm replaces all loops nested within the current loop block with \textit{markers} (lines 15 to 22), thereby generating the required snippets. 
These \textit{markers}, which can be seen in \autoref{fig:decompile_example}(b) and (c) as text highlighted in yellow, play a pivotal role in \tool{} approach. In \autoref{fig:decompile_example}(d), we demonstrate the successful reassembly of the \texttt{C} snippet using these \textit{markers}.

\subsubsection{\textbf{Temporal and Spatial Information Supplement.}}
\label{sec:information supplement}
To address the issue of nested loops, we employ the slicing strategies; however, this method results in the loss of informational integrity that is inherent in a complete function.
Relying solely on \texttt{wat} snippets as the entire input is insufficient. We provide necessary information from both temporal and spatial dimensions to enhance the input. With the aid of this contextual information, the LLM can achieve better performance.

From a \textbf{temporal} perspective, \texttt{wat} snippets are decompiled sequentially, and a snippet decompiled later does not know which local variables have already been defined in the previous ones. If an undefined variable is used in a snippet, it might lead to instability in the decompilation results. For example, the model might fabricate a result that appears to be correct or use a variable defined within the snippet as a substitute.

\textbf{Spatially}, \texttt{wat} may invoke functions whose declaration information is not present within the snippet. However, the number of arguments and return values for these functions are crucial and unknown, directly affecting the stack balance in Wasm. Thus, we also need to parse the function declarations for any functions called within the current snippet and incorporate this spatial information into the prompt as well.

\subsubsection{\textbf{Variable Renaming}}
\label{sec:rename_scheme}
Despite successfully slicing the \texttt{wat} code, we faced a significant hurdle during the reassembly of high-level language fragments: the variable names were randomized, with different segments potentially using different identifiers for what should have been the same variable. 

\begin{figure}[t]
    \centering
    \includegraphics[width=0.8\linewidth]{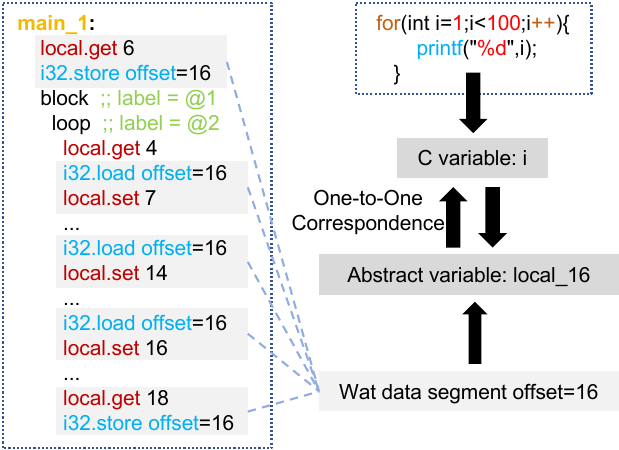}
    \caption{The mapping from the C variable to the abstract variable corresponding to the offset in \texttt{wat}.}
    \label{fig:Variable_Renaming_Scheme}
\end{figure}

As illustrated in \autoref{fig:Variable_Renaming_Scheme}, an analysis of the \texttt{wat} snippet reveals multiple \texttt{wat} local variables (identified as \texttt{i32\_6}, \texttt{i32\_7}, \texttt{i32\_14}, \texttt{i32\_16}, \texttt{i32\_18}) serving as carriers for values stored at the offset position of 16, with these values undergoing continuous changes. 
Consequently, we cannot find a variable entity in the \texttt{wat} code snippet that corresponds exactly to a variable in a high-level language.
In that case, we need to abstract a variable that can be directly correlated with the variable in a high-level language, ensuring that its naming is uniquely independent of the snippet's context. Our variable renaming scheme adopts the notation \texttt{local\_offset} to represent this abstract variable, which means that it is a local variable in the high-level language and stored at an offset in \texttt{wat}. 
Based on the DWARF debugging information\cite{Dwarf}, we establish a one-to-one correspondence between every variable name in the high-level function and an offset in \texttt{wat} function, thereby systematically renaming all variables in the source code.

\subsection{Modeling}
\label{sec:Modeling}
In this section, we delineate our approach to decompiling Wasm snippets using a code LLM. Our task is to translate \texttt{wat} snippets back into their corresponding \texttt{C} snippets. To tackle this challenge, we have fine-tuned a state-of-the-art, self-supervised code LLM that is adept at handling Wasm.

\subsubsection{\textbf{Model Architecture}}
We utilize CodeLlama-7b-hf~\cite{2024codellama} as the base model, inspired by its status as a leading code LLM and its exemplary performance in code-related tasks~\cite{azerbayev2023llemma,jin2023binary,wang2023mac}.
Given that the model architecture follows the standard transformer framework, we will not delve into a detailed review of this architecture. We now elaborate on how we adapted the model to our specific task objectives through a customized fine-tuning process, aimed at enhancing its decompilation effectiveness for Wasm.

\subsubsection{\textbf{Prompt Synthesis}}
Our prompt synthesis process utilizes all the features outlined in \autoref{tab:features}. The key distinction between fine-tuning prompt synthesis and evaluation prompt synthesis lies in the inclusion of the decompiled result, i.e., the renamed \texttt{C} snippet, at the end of the fine-tuning prompts. We now use the fine-tuning prompt as an example to illustrate our approach.

Given a \texttt{wat} text \( w = \{w_1, w_2, \ldots, w_{|w|}\} \), for any \texttt{wat} snippet \( w_n \), we need to capture temporal information, specifically the variables already defined up to that point, denoted as \( Vb_n = Vb_{n-1} \cup Vars_{n-1} \), where \( Vb_i \) represents the set of variables defined before the snippet and \( Vars_{n-1} \) represents the set of variables in the \( w_{n-1} \) snippet. We also need to gather spatial information, which refers to the function declarations \( d_n \) of the functions called within the current snippet. Consequently, our prompt \( p_n \) can be defined as:
\[
p_n = \{ [\text{Instruction}], i, [\text{Input}], Vb_n, d_n, w_n, [\text{Response}], c_n\}
\]
where, [\texttt{Instruction}], [\texttt{Input}], [\texttt{Response}] are labels; \( i \) represents the content of \textit{Instruction}, i.e., a series of pre-set constraints; \( vb_n, d_n, w_n \) constitute the \textit{Input} content, representing the already defined variables, the function declarations of the invoked functions, and the \texttt{wat} code, respectively; and \( c_n \) is the expected \textit{Response} content, which is the corresponding \texttt{C} code with renamed variables.

The \textit{Instruction} part should articulate the relationship and logic between the \textit{Input} and \textit{Response} to enhance the stability of the decompilation process. Firstly, it is essential to broadly describe the task the model is expected to perform, such as ``You are a \texttt{wat2c} model, tasked with decompiling the provided \texttt{wat} snippet into an equivalent \texttt{C} code snippet.'' 
Secondly, the decompiled C code should be logically and functionally equivalent to the \texttt{wat} code, with a strong emphasis on readability. Thirdly, since we have defined \textit{markers} as described in \S\ref{sec:slicing strategy}, it is crucial that these \textit{markers} are retained in the decompilation output. Lastly, we have specific naming conventions for the variables in the generated \texttt{C} code, which necessitates a clear directive to format the variable names as \texttt{local\_offset}. 
This structured approach in the \textit{Instruction} construction phase is pivotal for guiding the model to produce accurate and consistent decompilation results.

In the \textit{Input} part, it is crucial to provide sufficient yet non-redundant information, encompassing the main information and supplementary details. The main information is the \texttt{wat} snippet. Given its substantial length, we place this primary information at the end of this part. The supplementary information in our context pertains to temporal and spatial details, essentially the context of the \texttt{wat} snippet, including variables already defined before the current snippet and the function declarations of any functions called within the snippet. 
This comprehensive approach ensures that all necessary context and details are provided to facilitate an accurate and effective decompilation process.

\subsubsection{\textbf{Finetuning \& Interfering}}
We utilized PyTorch 2.2.1 and cuda 12.1 in Python to fine-tune and infer our model, setting the maximum sequence length to 2,048 for context. Our fine-tuning leveraged a LoRA dropout of 0.05 to mitigate overfitting and employed the CAUSAL\_LM task type, focusing on sequential token predictions. Optimization was achieved with a learning rate of 3e-4 and the adamw\_torch optimizer, ensuring efficient and effective model convergence.
During the fine-tuning process, we monitored the accuracy on the validation set to determine the convergence of our model. If convergence was observed, we implemented early stopping to prevent overfitting. Subsequently, we selected the best-performing model from the validation phase for our final evaluation, ensuring that we used the most effective iteration of our model for an accurate and reliable assessment.
All the experiments are run on a Linux server running Ubuntu 22.04.3 LTS with AMD EPYC 7713 64-Core Processor running at 516GB memory, and 2 Nvidia a100 80G GPUs.

\section{Evaluation}
\label{sec:evaluation}
In this section, we report and analyze the experimental results to address the following research questions (RQs):

$\bullet$ \textbf{RQ1:} Can \tool{} approach accurately decompile \texttt{wat} snippets into logically equivalent \texttt{C} snippets?

$\bullet$ \textbf{RQ2:} Do the decompiled results have high readability?

$\bullet$ \textbf{RQ3:} Can the results of our decompilation be recompiled and executed successfully?

$\bullet$ \textbf{RQ4:} How stable is the usage of the LLM for Wasm decompilation, and does it tend to generate hallucinations?

\subsection{Experimental Setup}

\textbf{Benchmark.} 
As mentioned in \S\ref{sec:dataset collection}, we compiled \texttt{C} files from \texttt{TheAlgorithms/C}~\cite{algorithm_c} and \texttt{bigcode/the-stack-dedup} dataset~\cite{Kocetkov2022TheStack} into Wasm binaries, which were then used to evaluate the performance of our approach. 
The evaluation dataset encompasses a total of 3,977 \texttt{C} files, providing a comprehensive foundation for analysis.

\noindent\textbf{Baselines.}
Due to the limited availability of tools for Wasm decompilation, we have identified only \texttt{Wasm2c}, \texttt{Wasm-decompile}, and \texttt{Ghidra}'s Wasm plugins as suitable for use.
The output of \texttt{Wasm-decompile} is not in \texttt{C}, as illustrated in \autoref{fig:decompile_example}(g), leading to its exclusion from the baselines.
Given the lengthy nature of Wasm, we couldn't identify any LLMs, including our base model, CodeLlama-7b-hf, that could decompile Wasm functions without specialized fine-tuning. Not even the top-performing GPT-4~\cite{openai2023gpt4} can manage this challenge effectively.
We thus exclude ML-based tools from our baselines.
Therefore, we finally select \texttt{Ghidra} and \texttt{Wasm2c} as our evaluation baselines.

\noindent\textbf{Evaluation Metrics.}
Due to the complexity of \texttt{C} language, the same syntactic and semantic expressions can manifest in different forms and structures. Furthermore, given the inherent unpredictability of LLM's outputs, the decompilation results by the \tool{} approach may be optimized related to the source code, such as by omitting intermediate variables, which can lead to differences in line numbers, complexity, and other factors. 
Consequently, assessing the code quality of decompilation results presents a significant challenge. Our evaluation is conducted based on the following criteria:

\underline{AST Edit Distance Similarity.}
To gauge structural consistency between the predicted and decompiled code, Song~\ea~\cite{song2024revisitingcodesimilarityevaluation} introduces AST edit distance similarity (AED-S). This metric is based on the AST edit distance (AED), which quantifies the minimal number of transformations required to convert one AST into another. AED-S offers a robust measure of structural similarity that transcends mere textual comparison and provides a normalized similarity score.

Building upon the AST edit distance, AED-S can be calculated using \autoref{formula:1}:
\begin{equation}
\text{AED-S}(T_1, T_2) = 1 - \frac{AED(T_1, T_2)}{\max(|T_1|, |T_2|)}
\label{formula:1}
\end{equation}
where $AED(T_1,T_2)$ is the edit distance between two ASTs, $|T_1|$ and $|T_2|$ represent the sizes (number of nodes) of the ASTs $T_1$ and $T_2$ respectively.
By normalizing the edit distance with respect to the size of the larger AST, AED-S provides a fair comparison between ASTs of different sizes.

\underline{Cyclomatic Complexity.}
To gauge the alterations in code complexity following decompilation, we employ cyclomatic complexity~\cite{1702388} as a key metric. This measure quantifies the count of linearly independent paths within the program, providing valuable perspectives on the maintainability of the code and highlighting any shifts in complexity that may have occurred during the decompilation process.
The cyclomatic complexity is calculated using \autoref{formula:2}:
\begin{equation}
V(G) = E - N + 2P
\label{formula:2}
\end{equation}
where \( V(G) \) represents the cyclomatic complexity, \( E \) denotes the number of edges within the control flow graph, \( N \) indicates the number of nodes in the control flow graph, and \( P \) is the number of connected components, which typically equals 1 for a standalone program.

\underline{Cosine Similarity.}
Cosine similarity is a token-based text similarity measure used to calculate the cosine of the angle between two non-zero vectors in a multidimensional space, reflecting their directional similarity. Compared to other token-based similarities like Jaccard~\cite{bag2019efficient}, cosine similarity considers the weight (frequency) in the vectors, which is particularly useful for this context as it thoroughly represents the frequency of occurrence of keywords and variable names.

\underline{CodeBLEU.}
CodeBLEU~\cite{ren2020codebleu} is an evaluation metric specifically designed for code generation tasks, particularly used in the fields of machine learning and natural language processing~\cite{le2022coderl,lu2021codexglue,wang2022no} to assess the quality of automatically generated source code. 
It builds upon the traditional BLEU~\cite{papineni2002bleu}, widely used in machine translation tasks to evaluate the similarity between reference and corresponding generated text. 
CodeBLEU recognizes the unique characteristics and requirements of programming languages, such as syntactic correctness, logical consistency, and data flow consistency. It offers a more comprehensive and accurate assessment of code generation, making it particularly suitable for evaluating the nuances of code that conventional text evaluation metrics might overlook.

\begin{table*}[t]
  \centering
  \caption{Evaluation results of the baselines and our \tool{} approach.}
  \resizebox{0.65\linewidth}{!}{
    \begin{tabular}{cccccccccc}
    \toprule
    \multicolumn{1}{c}{\multirow{2}[4]{*}{\textbf{Decompilation tools}}} & \multicolumn{3}{c}{Similarity} & \multicolumn{2}{c}{Accurancy} & \multicolumn{2}{c}{Readability} \\
\cmidrule(r){2-4} \cmidrule(rl){5-6} \cmidrule(l){7-8}            & AED-S   & CCN   & COS   & \multicolumn{1}{l}{CodeBLEU} & \multicolumn{1}{l}{C@Func} & Bloat Rate & C@Syntax \\
    \midrule
    Wasm2c  & —     & 0.5413 & 0.5496 & —     & — & 964.84\% & —  \\
    \midrule
    Ghidra & 0.2565 & 0.6687 & 0.6915 & 0.2589 & 0.6403 & 116.94\% & 0.4997 \\
    \midrule
    \textbf{\tool{}}  & \textbf{0.7293} & \textbf{0.7237}  & \textbf{0.9724} & \textbf{0.6353} & \textbf{0.8829} & \textbf{3.34\%} & \textbf{0.8908}  \\
    \bottomrule
    \end{tabular}%
    }
  \label{tab:compare_with_metrics}%
\end{table*}%

\underline{Code Bloat.}
One of the most intuitive indicators of code readability is the change in the number of lines of code~\cite{5332232}. We assume that the readability of the source code is relatively high, and recovering the source code is a primary goal of decompilation. Excessive bloat dilutes original semantics, making code harder to understand. Therefore, introducing code bloat, the rate of expansion in the number of code lines, as a metric is a wise choice.

\underline{Syntactic Completeness \& Function Completeness.} Syntactic completeness is a foundational metric for the legitimacy of code syntax and is widely adopted in the field. 
This metric measures the proportion of syntactically complete code relative to the total codebase, with the help of \texttt{tree-sitter}~\cite{Tree_sitter}, which can determine whether a piece of code or a function has errors.
Additionally, syntactic completeness underpins compilability; code that is not syntactically complete is inherently uncompileable.
Function completeness refers to whether the decompiled code includes all the functions from the source code.

\underline{Recompilation \& Re-execution.}
Tan~\ea~\cite{tan2024llm4decompile} introduce recompilability and re-executability as metrics to evaluate the success of decompilation.
Codes are fragile, and even minor modifications can lead to syntax violations, resulting in compilation or execution errors. When decompiled code can be recompiled, it provides strong evidence of its syntactic integrity. This ensures that the decompiled code adheres to the structural and syntactic standards expected by compilers. When decompiled code can be re-executed, we assess its semantic consistency by comparing the outputs.

\subsection{RQ1: Similarity \& Accuracy}
To evaluate the accuracy of the decompiled results, we examined two aspects: (1) We assessed the similarity between the decompiled code and the source code. Specifically, we employed three similarity metrics: AST edit distance similarity, cyclomatic complexity, and cosine similarity, which respectively describe the consistency between the decompiled results and the source code in terms of node structure, complexity, and symbolic representation, jointly reflecting the logical similarity. (2) We also investigated function completeness, quantifying the degree to which functions are preserved in the decompiled results. Specifically, function completeness is measured by calculating the proportion of functions in the decompiled code to those in the source code.

These metrics employed in our study utilize classic algorithms or Python packages, with certain customizations made to suit our analysis needs. 
For the AST edit distance similarity (AED-S), we adopt the standard Levenshtein distance algorithm by using method from \texttt{apted}\footnote{\url{https://github.com/JoaoFelipe/apted}.} Python package. To ensure that the properties depicted by the AED-S and token-based cosine similarity do not overlap, we exclusively utilize information about AST node types, omitting the text associated with the nodes.
The cyclomatic complexity is measured using the \texttt{lizard}\footnote{\url{https://github.com/terryyin/lizard}.} Python package. To manage the excessively long decompilation results from \texttt{Ghidra} and \texttt{Wasm2c}, and to eliminate unnecessary library functions, we format the decompilation outputs, retaining only the necessary functions.
Cosine similarity is calculated using methods from the \texttt{sklearn}\footnote{\url{https://pypi.org/project/sklearn}.} Python package to ensure standardized results. 
Additionally, we rename all variables in the source code according to the variable renaming scheme described in \S\ref{sec:rename_scheme}, to minimize the impact of variable name discrepancies on the cosine similarity and variable completeness.
The reason for disregarding \texttt{Wasm2c}'s AED-S is due to its excessively lengthy and redundant decompilation output, which is fundamentally similar to \texttt{wat} code, rendering the comparison meaningless.

\textbf{Overall Results.}
The results presented in \autoref{tab:compare_with_metrics} demonstrate that \tool{} outperforms existing Wasm decompilation tools regarding similarity and accuracy. 
Looking at similarity metrics, \tool{} surpasses the state-of-the-art tools across three indicators: AST edit distance similarity (AED-S), cyclomatic complexity (CCN), and cosine similarity (COS). Specifically, \tool{} achieved an AED-S of 0.7293, 1.85 times higher than that of \texttt{Ghidra}, indicating a high structural similarity of the AST to the source code. The CCN reached 0.7237, an improvement of 8\%, which underscores \tool{}'s ability to accurately reconstruct complex loop structures—a known theoretical advantage of traditional tools such as \texttt{Wasm2c} and \texttt{Ghidra}. The COS score of 0.9724, a 41\% increase, suggests superior restoration capabilities at the token level, including keywords and variable names.
From an accuracy perspective, \tool{} demonstrates outstanding performance on the CodeBLEU and function completeness (C@Func), recording scores of 0.6353 and 0.8829, respectively. These scores substantially surpass the baselines, which only achieved 0.2589 and 0.6403 in these metrics, representing improvements of 145\% and 37\%, respectively. This indicates that \tool{} not only excels in recovering the syntax and structure of the decompiled code but also ensures a high degree of functional fidelity, making it a superior choice for practical applications in Wasm decompilation.

It is worth noting that theoretically, each function should produce an output; however, the function completeness (C@Func) metric does not equal 1. This discrepancy arises due to the absence of a particular function in the decompiled code. When processing decompilation results, if \texttt{tree-sitter} identifies a function as containing errors, we remove these syntactically incorrect functions. Subsequently, an additional result file is generated, which is then specifically used to calculate C@Func.

\subsection{RQ2: Readability}
Code readability is fundamentally determined by two main factors: Syntactic Completeness and Code Bloat. Syntactic completeness can be effectively assessed using \texttt{tree-sitter}, which not only can verify if it has an error but also if something is missing. On the other hand, Code Bloat is quantified by directly comparing the number of lines between the source code and the decompiled output. 
It is important to note that a more precise measure is the absolute rate of change in the number of lines, with an emphasis on minimizing this value. This is based on the assumption that the readability of the source code is relatively high.

Due to the excessively lengthy and redundant decompilation results from \texttt{Ghidra} and \texttt{Wasm2c}, which include numerous unnecessary library functions, we performed preprocessing on these outputs. This involved extracting the source code functions and removing comments while standardizing the format. Additionally, we incorporated other non-function elements from the source code to ensure a more comprehensive and cleaner representation in our analysis.

\textbf{Overall Results.}
Our results demonstrate that \tool{} achieves the lowest code bloat rate at just 3.34\%, significantly lower than the baselines and close to the source code level. We observed that \tool{} tends to merge the initialization of variables of the same type and retains only variables with corresponding concrete storage addresses in Wasm, eliminating redundant intermediate variables. This approach results in a more compact and readable decompiled output. Additionally, \tool{} achieves a syntax completeness rate of 0.8906\%, which is 1.78 times higher than the state-of-the-art tools. This indicates that \tool{} has a robust understanding and application of Wasm and \texttt{C} syntax, producing compliant and highly readable decompiled results.

\subsection{RQ3: Reusability}

The evaluation methods for reusability, namely recompilability and re-executability, are straightforward: compile the decompiled \texttt{C} code into Wasm, and then compare the output consistency between the original Wasm file and the recompiled file. However, recompilability and re-executability have always been challenging areas. As Liu~\ea\cite{liu2020far} noted, the academic community generally holds a conservative and pessimistic view on the correctness of decompilation. Specifically, in the field of Wasm decompilation, Wasm's unique linear memory and stack structure further complicate its correctness, thereby affecting its recompilability and re-executability. In this experiment, we excluded \texttt{Wasm2c}, as it primarily performs a line-by-line translation from \texttt{wat}, resulting in \textbf{excessive code bloat} (as shown in \autoref{tab:compare_with_metrics}), which makes the discussion about its recompilability meaningless.

\begin{table}[htbp]
  \centering
  \caption{Evaluation results of the reusability of \tool{}.}
    \begin{tabular}{c|ccc}
    \toprule
    \multicolumn{1}{l|}{Reusability} & \multicolumn{1}{l}{Recompilation} & \multicolumn{1}{l}{Re-execution} & \multicolumn{1}{l}{Consistency} \\
    \midrule
    \texttt{Ghidra} & 0.90\%  & 0.90\%  &0.89\%\\
    \textbf{\tool{}} & \textbf{52.11\%} & \textbf{43.55\%} & \textbf{27.15\%} \\
    \bottomrule
    \end{tabular}%
  \label{tab:Reusability}%
\end{table}%

\textbf{Overall Results.}
In our experiments, we found that the code decompiled by \texttt{Ghidra}'s Wasm plugin \textbf{could not be directly compiled}. This issue arises not only because the main function has an alias but also because \texttt{Ghidra} fails to recover strings, instead using placeholders. Unfortunately, these placeholders are not defined, leading to compilation errors.
Furthermore, we observed frequent erroneous annotations during the decompilation process, a phenomenon common across both Windows and Ubuntu environments. This highlights a substantial loss of information in \texttt{Ghidra}'s conversion from Wasm to \texttt{C}. In sharp contrast, \tool{} approach achieved a recompilability rate of 52.11\%, a re-executability consistency of 43.55\%, and an output consistency rate of 27.15, significantly surpassing \texttt{Ghidra}, which scored only around 0.9\% in all three metrics. Moreover, the few cases where successful compilation and execution occurred with \texttt{Ghidra} were exclusively due to the absence of strings and the extreme simplicity of the code. This demonstrates that \tool{} not only possesses robust understanding and application capabilities of both Wasm and \texttt{C} syntax but also effectively interprets high-level semantic information, significantly enhancing the readability of the decompiled code.

\subsection{RQ4: Robustness}
The robustness of outputs from LLMs remains a significant challenge. Despite fine-tuning, some patterns from the pre-training of LLMs are deeply ingrained and influence subsequent outputs. Robustness is multifaceted.

\textbf{Stability.}
The stability of outputs from LLMs has often been a subject of skepticism, which is justifiable given that untrained LLMs indeed exhibit such issues. To measure the impact of the factor, we experimented by randomly selecting ten syntactically complete cases and performing multiple repetitive experiments. Specifically, we conducted ten consecutive decompilations for each case and calculated the CodeBLEU score, computing the variance for each before averaging the results. Surprisingly, \textbf{the variance for each case was zero}, indicating that the CodeBLEU results were consistent across all trials. This conclusively demonstrates the stability of \tool{}. Additionally, we considered the issue of output formatting, and the experiments showed that \tool{}'s outputs consistently adhered to the predefined format, as also indicated by the variance results.

\textbf{Hallucination.}
In terms of hallucinations, although \tool{} may fabricate some variable names, these variables accurately represent the real Wasm addresses and their naming conventions align with the standards we have established. A typical scenario arises when dealing with string variables; often, these strings are not found in the data segment but are instead constructed in the function code as arrays of numbers located at consecutive addresses. Consequently, we encounter variables in the decompiled results that are not visible in the source code.

\section{Discusstions}
\label{sec:limitations}

\subsection{Case Study}

\begin{figure}[ht]
    \centering    \includegraphics[width=\linewidth]{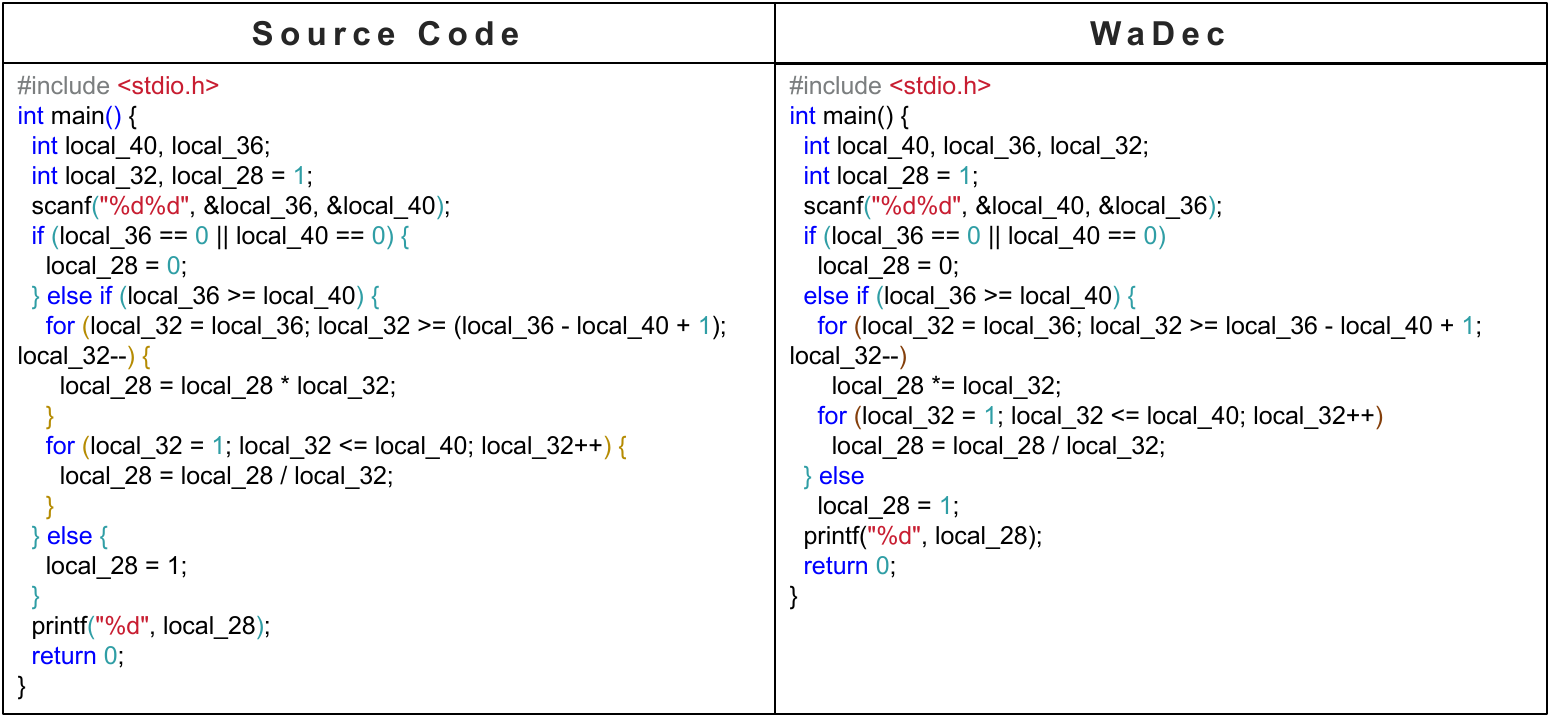}
    \caption{A concrete example of \tool{}'s performance and the related source code snippet (with renamed variables), which is from the \texttt{bigcode/the-stack-dedup} dataset.}
    \label{fig:case_study_wadec}
\end{figure}

\begin{figure}[ht]
    \centering    \includegraphics[width=\linewidth]{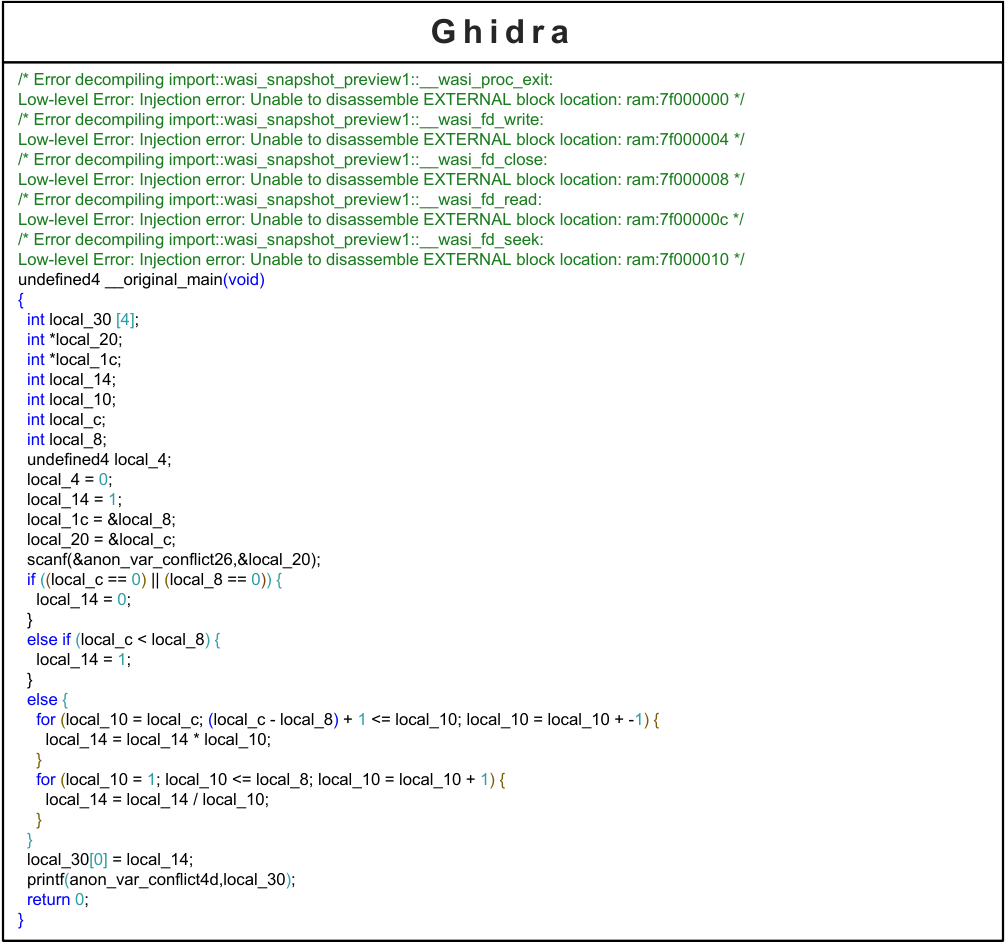}
    \caption{A concrete example of \texttt{Ghidra}'s performance.}

    \label{fig:case_study_ghidra}
\end{figure}

To vividly demonstrate the superiority of \tool{}, we provide concrete examples, as shown in \autoref{fig:case_study_wadec} and \autoref{fig:case_study_ghidra}.
We can see apparent issues with the decompilation result of the Wasm plugin in \texttt{Ghidra}. 
First, there is \textbf{error information} in the decompilation result, which clearly indicates that \texttt{Ghidra} has certain problems with decompiling Wasm. 
Second, by comparing \texttt{Ghidra}'s decompilation result with the source code, we can find that \texttt{Ghidra} cannot recover \textbf{string constants}, which severely affects the semantics and I/O of the code. 
Moreover, compared with the source code and \tool{}'s output, there is a significant\textbf{ code bloat}. 
Finally, there are varying degrees of \textbf{inconsistencies} in function names, variable types, and other aspects.
We can observe that \tool{} outperforms \texttt{Ghidra} regarding similarity, accuracy, readability, and reusability.


\subsection{User Study}
We conducted a user study with 20 participants to compare the decompilation capabilities of \texttt{Ghidra} and \tool{}. Evaluation criteria included code readability and accuracy. Out of 100 randomly selected pairs, participants preferred \texttt{Ghidra} in only two cases, attributable to LLM-generated hallucinations rather than superior readability.
Additionally, we surveyed participants on the necessity of effective Wasm decompilers. All 5 Wasm-familiar participants, including 2 experts, deemed highly readable decompilation extremely necessary. Among 10 participants aware of but inexperienced with Wasm, 8 considered decompilation necessary or extremely necessary.

\subsection{Limitations}

\noindent\underline{\textbf{Limited model selection.}}
Due to resource constraints in our experiments, we were unable to test a broader range of LLMs, as their fine-tuning and inference require substantial GPU memory and time. Therefore, we narrowed our model selection to LLMs specifically pre-trained for coding tasks. CodeLlama, a model enriched with Wasm knowledge, is tailored for handling longer stretches of code and supports longer prompt inputs, making it ideal for extensive code tasks. Our findings indicate that CodeLlama, when used as the foundational model for fine-tuning our tool, delivers superior performance, outstripping state-of-the-art tools.

\noindent\underline{\textbf{Relatively lower CodeBLEU scores.}}
Compared to other work on general binary (not for Wasm) decompilation~\cite{tan2024llm4decompile}, our results using the same CodeBLEU metric are relatively lower. To understand this discrepancy, we conducted a manual review. We discovered that, due to the rich semantics of the \texttt{C} language, code with identical logic and functionality can have different representations. 
Furthermore, we have pinpointed several limitations inherent in CodeBLEU when it comes to accurately assessing the code quality in certain scenarios. As a result, we are advocating for the development of more refined evaluation metrics by the research community.
We have provided an in-depth exploration of these findings in our open-source artifact~\cite{wadec}, where we present several concrete examples that shed light on the intricacies of the phenomenon.

\noindent\underline{\textbf{Limited structure and longer time.}}
We observed that due to the unique properties of Wasm, \tool{} struggles with data structures, often treating them merely as multiple variables or arrays. This limitation makes it somewhat complementary to traditional techniques. Combining them could potentially yield more refined results. 
Additionally, \tool{} typically requires more time to decompile once, and given the lengthy nature of \texttt{wat} code and our slicing strategy, a function may need to be decompiled multiple times, significantly reducing efficiency.
For shorte prompts, it takes about 0.1 seconds each, while prompts approaching the LLM's input limit may take 1-2 seconds each to complete. 
Therefore, accelerating the decompilation rate of LLMs presents a worthwhile challenge and could greatly advance the state of the art in the field of decompilation.

\noindent\underline{\textbf{Impact of optimization level.}}
The \texttt{O1} optimization includes techniques like constant folding, variable reuse, and direct integration of simple functions into their calling functions. The \texttt{O2} optimization level goes further by implementing more complex code restructuring, function inlining, and loop optimization.
Currently, our approach has not been adapted to these specific characteristics, nor have we utilized a dataset involving compilation optimization to fine-tune the LLM.
We thus don't evaluate the performance of \tool{} across various optimization levels. 
Fortunately, only 28\% of Wasm binaries on the web are minified~\cite{hilbig2021empirical}, which means that our work is effective for the majority of scenarios.
Furthermore, our approach is adaptable and can be customized to fine-tune LLMs for decompiling Wasm optimized at various levels. We intend to explore this further in future work.

\section{Related Work}
\label{sec:related work}

\textbf{Traditional Wasm analysis.}
Since the advent of Wasm~\cite{wasm_web}, there have been numerous studies focused on the security~\cite{Romano_Liu_Kwon_Wang_2021,Hilbig_Lehmann_Pradel_2021,johnson2023wave,chen2022wasai}, performance~\cite{Jangda_Powers_Berger_Guha_2019}, and application of Wasm across various scenarios~\cite{EOSIO,NEAR,Li_Dong_Gao_2021}. Analysis and testing techniques have encompassed static and dynamic analysis, as well as fuzz testing, among others. However, these analytical and testing methodologies do not directly yield results that are easily readable and understandable. Even with the support of official tools like Wabt~\cite{Wabt} and the assistance of the \texttt{wat} format for readability, developers are still required to interpret the high-level semantics behind each instruction. 
This interpretation process remains a significant challenge, underscoring the need for more intuitive tools and methods to bridge the gap between low-level instructions and high-level understanding.

\noindent \textbf{LLMs for decompilation.}
LLMs have demonstrated impressive performance in code generation tasks~\cite{Wang_Wang_Joty_Hoi_2021, KC_Morrison_2023}, with numerous studies utilizing these models to validate their feasibility in decompilation work~\cite{tan2024llm4decompile,al2023extending, Cao_Liang_Chen_Hu_2022}.
In recent years, large-scale code models such as CodeLlama~\cite{rozire2024code} and DeepSeek~\cite{lu2024deepseekvl}, along with their fine-tuned variants, have been introduced. Research on these large code models is becoming increasingly vibrant, inspiring researchers to delve deeper and invest more extensively in related fields.
For instance, Tan~\ea\cite{tan2024llm4decompile} utilized DeepSeek to fine-tune large-scale decompilation models ranging from 1B to 33B parameters, marking the first instance of an LLM centered on open-source decompilation.

\section{Conclusion}
\label{sec:conclusion}

In this study, we introduced \tool{}, an approach leveraging a fine-tuned LLM to decompile Wasm binary. The model's training on a specialized \texttt{wat-c} dataset, coupled with self-supervised learning, has proven pivotal in enhancing decompilation efficacy. Our results indicate that \tool{} significantly outperforms existing tools, achieving a minimal code inflation rate and maintaining high recompilability and re-execution rates. This advancement not only bolsters the readability and analyzability of Wasm code but also paves the way for more robust automated code analysis, optimization, and security auditing processes.

\section*{Acknowledgement}
This work was supported by the National NSF of China (grants No.62072046, 62302181), the Knowledge Innovation Program of Wuhan-Basic Research (2022010801010083), the Key R\&D Program of Hubei Province~(2023BAB017, 2023BAB079), and HUSTCSE-FiberHome Joint Research Center for Network Security.

\bibliographystyle{ACM-Reference-Format}
\bibliography{main}

\end{document}